\documentclass[twocolumn,showpacs,amsmath,amssymb,pre,aps]{revtex4}   
\usepackage{graphicx}
\usepackage{dcolumn}
\usepackage{bm}
\usepackage{color}

\renewcommand{\vec}[1]{\mathbf{#1}}

\newif\ifgraph

\graphtrue             %

\begin{document}
\title{
Diffusion of Chiral Janus Particles in Convection Rolls}

\author{Yunyun Li$^{1}$, Lihua Li$^{1}$,  Fabio Marchesoni$^{1,2}$, Debajyoti Debnath$^{3}$, and Pulak K. Ghosh$^{3}$\footnote{E-mail: pulak.chem@presiuniv.ac.in (corresponding author)}}
 \affiliation{$^{1}$ Center for Phononics and Thermal Energy Science, Shanghai
 Key Laboratory of Special Artificial Microstructure Materials
 and Technology, School of Physics Science and Engineering, Tongji University, Shanghai 200092, China}
 \affiliation{$^{2}$ Dipartimento di Fisica, Universit\`{a} di Camerino, I-62032 Camerino, Italy}
 \affiliation{$^{3}$ Department of Chemistry,
Presidency University, Kolkata 700073, India}

\date{\today}

\begin{abstract}
The diffusion of an artificial active particle in a two-dimensional
periodic pattern of stationary convection cells is investigated by means of
extensive numerical simulations. In the limit of large P\'eclet numbers,
i.e., for self-propulsion speeds below a certain depinning threshold and
weak roto-translational fluctuations, the particle undergoes asymptotic
normal diffusion with diffusion constant proportional to the square root of
its diffusion constant at zero flow. Chirality effects in the propulsion
mechanism, modeled here by a tunable applied torque, favors particle's
jumping between adjacent convection rolls. Roll jumping is signaled by an
excess diffusion peak, which appears to separate two distinct active
diffusion regimes for low and high chirality. A qualitative interpretation
of our simulation results is proposed as a first step toward a fully
analytical study of this phenomenon.

\end{abstract}
\maketitle

\section{Introduction} \label{intro}

Microswimmers are Brownian particles capable of self-propulsion
\cite{Granick,Muller}. The simplest category among them consists of
artificial micro- and nano-propellers, which, due to some {\it ad hoc}
asymmetry of their geometry and/or chemical composition, are capable of
harvesting environmental energy and convert it into kinetic energy. The
artificial microswimmers most investigated in the literature are the
so-called Janus particles (JP), basically spherical colloidal particles with
two differently coated hemispheres, or ``faces''. Their axial propulsion is
sustained by the dipolar near-flow-field they generate by interacting with
the surrounding active (mostly highly viscous) medium
\cite{Marchetti,Gompper}.

\begin{figure}[tp]
\centering \includegraphics[width=8.5cm]{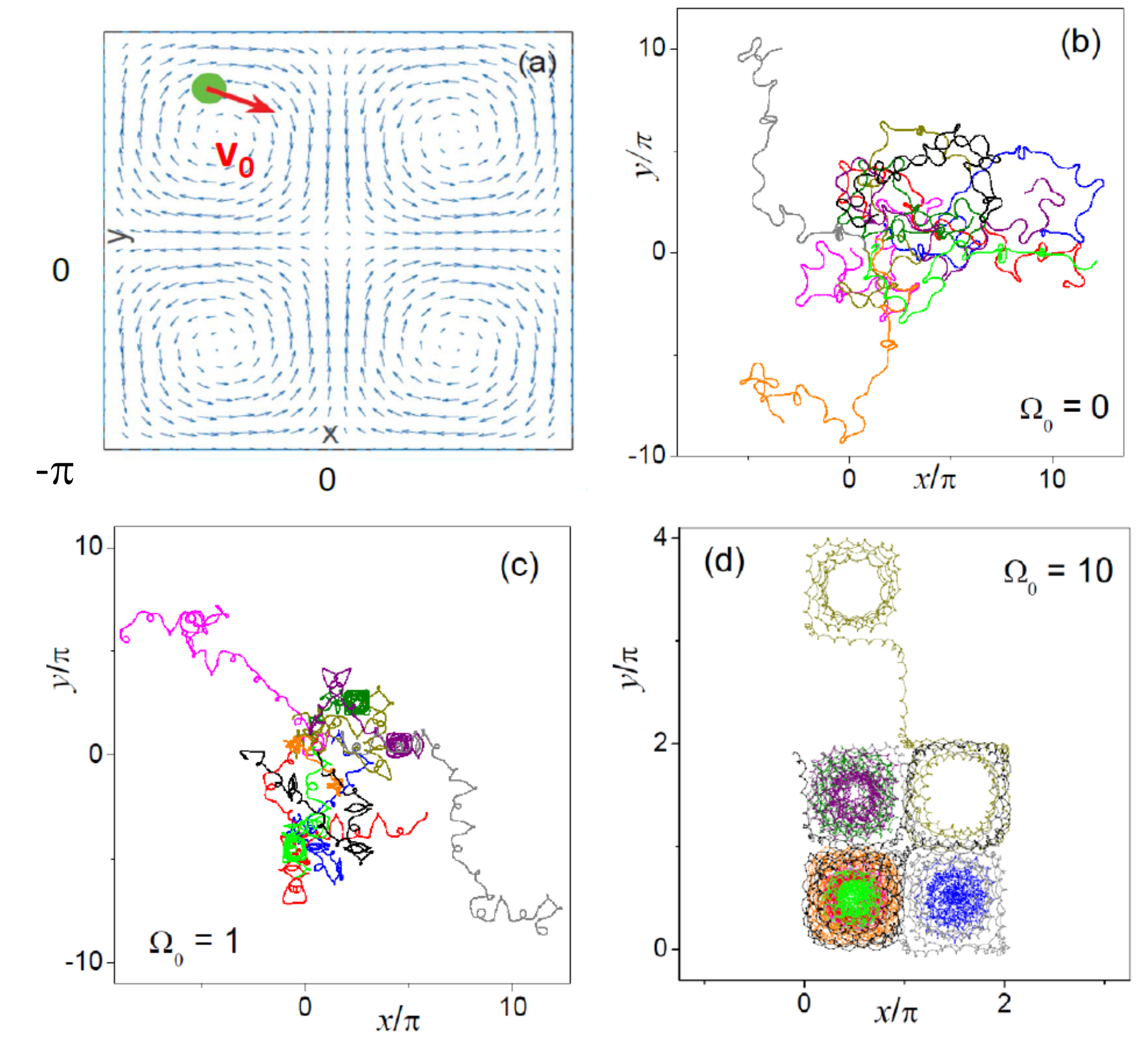}
\caption{Diffusion of an active JP in a 2D periodic pattern of convective
stationary rolls. (a) Unit flow cell, ${\vec v}_\psi =(\partial_y,
-\partial_x)\psi$, with $\psi(x,y)$ of Eq. (\ref{psi}), consisting of four
counter-rotating subcells. (b)-(d) Trajectory
samples of length $t$=100, respectively for $\Omega_0=0$ (achiral), $1$ (depolarized),
and $10$ (highly chiral). Other model parameters are: $\alpha=1$,  $U_0=v_0=1$,
$D_0=D_\theta=0.003$ and $L=2\pi$. Note that for $v_0<v_{\rm th}$,
roll jumps are noise activated (see text). \label{F1}}
\end{figure}

Recently, artificial microswimmers found promising applications in the
pharmaceutical  (e.g., smart drug delivery \cite{smart}) and medical research
(e.g., robotic microsurgery \cite{Wang}), whereby one expects that the
function they are designed to perform is governed in time and space by their
diffusive properties. To this regard it is important to control the diffusion
of active particles in crowded \cite{Gompper} and patterned environments
\cite{Bechinger}, where they interact with other system components, either
chemically \cite{aDLR} or mechanically \cite{stirrer}. Even more important
for applications to cell biology and chemical industry is regulating their
diffusion in hydrodynamically active mediums \cite{Marchetti,Lauga}

To this purpose we investigated the diffusion of a single overdamped JP with self-propulsion
speed $v_0$ suspended in a two-dimensional (2D) stationary laminar flow with periodic
center-symmetric stream function
\begin{equation}
\label{psi}
\psi(x,y)= ({U_0L}/{2\pi})\sin({2\pi x}/{L})\sin({2\pi y}/{L}),
\end{equation}
where $U_0$ is the maximum advection speed and $L$ the wavelength of the flow
unit cell. As illustrated in Fig. \ref{F1}, $\psi(x,y)$ defines four
counter-rotating advection subcells, or, following the notation of Ref.
\cite{Pomeau}, convection rolls. Particle transport in such a flow pattern
has been studied under diverse conditions. For instance, in the presence of
periodic perturbations the {\em deterministic} dynamics of a {\it passive}
particle exhibits remarkable chaotic properties \cite{Gollub1,Gollub2}.
Especially relevant to the present work are the results reported for the
diffusivity \cite{Pomeau} and the nonlinear mobility
\cite{Saintillan,Vulpiani} of {\em passive} tracers subject to thermal
fluctuations. Note that in Ref. \cite{Vulpiani} the drive acting on the
tracer plays the role of a self-propulsion velocity with fixed orientation.
However, despite its practical implications, the problem of how a flow field
with stream function like $\psi(x,y)$ can affect the diffusion of
self-propelled particles has not been fully investigated, yet. The problem
was addressed, indeed, by the authors of Ref. \cite{Neufeld}, but only in the
{\it noiseless}, chaotic limit. These authors also proved that, for
self-propulsion speeds below a certain threshold, the particle gets
dynamically trapped inside the convective rolls and its diffusion suppressed.

In this paper we consider the more realistic situation of an active particle
subject to both translational and orientational fluctuations. As a
consequence, the direction of its self-propulsion velocity is driven not only
by the local flow shear \cite{Neufeld} and, possibly, a chiral (applied or
intrinsic) torque \cite{Lowen}, but also by an intrinsic rotational noise.
Moreover, due to thermal fluctuations, random hopping between convection
rolls \cite{Pomeau} can occur even for self-propulsion speeds below the
trapping threshold of Ref. \cite{Neufeld}. As a result, active diffusion in
the laminar flow of Eq. (\ref{psi}) develops two distinct regimes,
respectively for low and high chirality, both with a peculiar dependence on
the particle's self-propulsion parameters. At the transition, the chiral and
shear torque compensate each other inside two diagonally opposite $\psi(x,y)$
subcells; this causes a partial depinning of the active particle from the
convection rolls with a consequent diffusivity surge.

\section{The model} \label{model}

In the plane $(x,y)$ the overdamped dynamics of an active JP can be
formulated by means of two translational and one rotational Langevin equation
(LE)
\begin{eqnarray} \label{LE}
\dot {\vec r}&=& {\vec v}_\psi + {\vec v}_0 +\sqrt{D_0}~{\bm \xi}(t) \\ \nonumber
\dot \theta &=& \Omega_0 + ({\alpha}/{2})~\nabla \times {\vec v}_\psi +\sqrt{D_\theta}~\xi_\theta (t),
\end{eqnarray}
where ${\vec r}=(x,y,)$, ${\vec v}_\psi =(\partial_y, -\partial_x)\psi$ is the advection
velocity and the self-propulsion vector, ${\vec v}_0=v_0(\cos \theta,
\sin \theta)$, has constant modulus, $v_0$, and is oriented at an angle
$\theta$ with respect to the $x$-axis.
The translational noises in the $x$ and $y$ directions, ${\bm
\xi}(t)=(\xi_x(t), \xi_y(t))$, and the rotational noise, $\xi_\theta (t)$,
are stationary, independent, delta-correlated Gaussian noises, $\langle
\xi_i(t)\xi_j(0)\rangle = 2 \delta_{ij}\delta (t)$ with $i,j=x,y,\theta$. As
long as diffusion takes place away from boundaries or other obstacles, the
particle can be taken as pointlike. Effects due to its actual geometry and
chemical-physical characteristics are encoded in the dynamical parameters
appearing in Eq. (\ref{LE}). $D_0$ and $D_\theta$ are the respective noise
strengths, which we assume to be unrelated for generality (e.g., to account
for different self-propulsion mechanisms \cite{ourPRL}). The reciprocal of
$D_\theta$ is the correlation, or angular persistence time of ${\vec v}_0$;
accordingly, $v_0/D_\theta$ quantifies the persistence length of the
particle's self-propelled random motion. The flow shear exerts a torque on
the active particle with frequency proportional to the local fluid vorticity
$\nabla \times {\vec v}_\psi$ \cite{selfpol1,selfpol2,selfpol3}. The constant
$\alpha$ can depend, in principle, on the properties of the particle's
surface and its fabrication process. Here, for simplicity, we adopt Fax\'en's
second law, which, for a spherical particle, yields $\alpha=1$
\cite{selfpol1,selfpol2}. For $\alpha=1$ and the stream function of Eq.
(\ref{psi}), the self-polarization term in the second LE (\ref{LE}) can be
conveniently rewritten as $(-1/2)\nabla^2\psi$ or $(2\pi/L)^2\psi(x,y)$; its
modulus is maximum, $\Omega_l$, at the center of each subcell. Finally,
$\Omega_0$ represents a constant torque, either applied by the experimenter
\cite{Sen_mag}, or intrinsic to the JP design \cite{Ghosh2,Wurger}, or
exerted by a bounding surface \cite{Stark}, or possibly due to unavoidable
fabrication defects \cite{Wang}. In any case, $\Omega_0$ is a measure of the
particle's chirality, which, as proven below, greatly impacts its diffusion.
Due to the symmetry of the LE (\ref{LE}) we restrict our analysis to the
domain $\Omega_0\geq 0$.

The LEs (\ref{LE}) can be conveniently reformulated in dimensionless units by
rescaling $(x,y) \to (\tilde x, \tilde y)=(2\pi/L)(x,y)$ and $t \to \tilde t=
\Omega_L t$ with $\Omega_L=2 \pi U_0/L$. Accordingly, the four remaining
independent parameters get rescaled as $v_0 \to v_0/U_0$, $\Omega_0 \to
\Omega_0/\Omega_L$, $D_0 \to D_0/D_L$ and $D_\theta \to D_\theta/\Omega_L$,
with $D_L=U_0L/2\pi$. This means that, without loss of generality, we can set
$L=2\pi$ and $U_0=1$ and the simulation results thus obtained can be regarded
as expressed in dimensionless units and easily scaled back to arbitrary
dimensional units. The stochastic differential Eqs. (\ref{LE}) were
numerically integrated by means of a standard Milstein scheme \cite{Kloeden}.
Particular caution was exerted when computing  the asymptotic diffusion
constant $$D=\lim_{t\to \infty} \langle [x(t) -x(0)]^2\rangle /2t.$$  Indeed,
upon lowering the noise strengths $D_0$ and $D_\theta$, the intercellular
diffusion of a trapped active JP gets suppressed; accordingly, the time
transients  grow exceedingly long.

We conclude this section with an important remark. With the term particle
trapping, we refer to the dynamical trapping caused by advection, which drags
the suspended particle along closed orbits. This phenomenon is not to be
mistaken with the trapping by an external potential. To this regard, we
suggest the reader to compare the problem at hand with the problem of active
diffusion in a planar ``egg carton'' potential \cite{Lammert} or in a square
array of truncated harmonic traps \cite{Misko}. In these two cases, the
underlying diffusion process is controlled by thermal activation, whereas in
the present problem  a crucial role is played by advection.

\section{Results} \label{results}

In this paper we focus on the phenomenon of advection dominated active
diffusion, that is on the dynamical regime where a {\it noiseless} achiral JP
would be strictly localized by the stream function $\psi(x,y)$. Indeed,
depinning of a noiseless particle from the dynamical trap represented by a
single convection roll occurs for self-propulsion speeds above a certain
threshold \cite{Neufeld}. For a qualitative estimate of such a depinning
speed, $v_{\rm th}$, we notice that a trapped active particle can only
perform circular orbits with radius, $v_0/\Omega_L$, not exceeding the
effective half-width of the convection roll, $R_{\rm s}$.  For a square
subcell of $\psi(x,y)$, Eq. (\ref{psi}), $R_{\rm s} \simeq L/2\sqrt{2}$;
hence, the trapping condition $v_0 <v_{\rm th}=\Omega_L R_{\rm s}$. For the
flow parameters used here, $v_{\rm th}\simeq 2.2$, in close agreement with
the numerical result obtained in Ref. \cite{Neufeld}.

\begin{figure}[tp]
\centering \includegraphics[width=8.5cm]{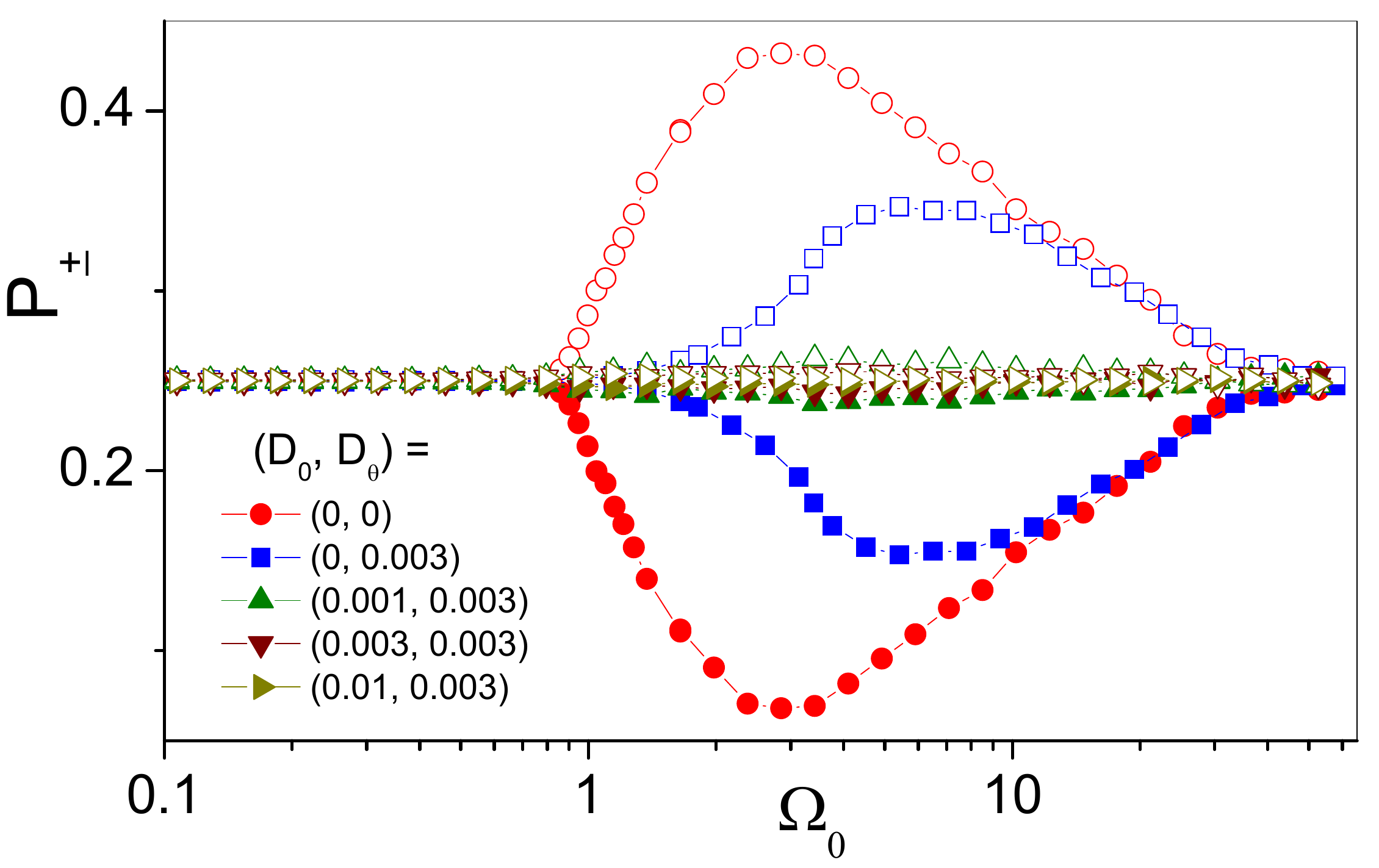}
\caption{Particle's probability density integrated respectively over a positive
($P_+$, solid symbols) and a negative subcell ($P_-$, empty symbols), vs. $\Omega_0$ for different
values of the noise strengths, $D_0$ and $D_\theta$ (see legend).
Other model parameters are: $\alpha=1$, $U_0=v_0=1$, and $L=2\pi$. Particle's depinning
from the negative subcells occurs for $\Omega_0 \sim \Omega_L$  \label{F2}}
\end{figure}

\subsection{Convection Rolls as Dynamical Traps} \label{traps}

The laminar flow exerts opposite shear (or self-polarization) torques in each
pair of neighboring $\psi(x,y)$ subcells. Accordingly, we term the convection
rolls positive or negative, depending on the sign of the particle's
self-polarization. As all simulation data reported here are for $\alpha=1$,
vorticity and self-polarization have the same sign -- positive in the
subcells centered at $(\pm L/2)(1,1)$ and negative in the subcells centered
at $(\pm L/2)(1,-1)$. Of course, the chiral torque, $\Omega_0$, in the second
Eq. (\ref{LE}) has a different impact on the particle's dynamics, depending
on the sign of the subcell considered. In a negative subcell, a positive
chiral torque with $\Omega_0\simeq\Omega_L$ tends to annul the
self-polarization torque. As a result, the ``depolarized'' particle can
escape the subcell even for $v_0 < v_{\rm th}$ and, in the absence of noise,
to sojourn inside one of the positive subcells. This mechanism is termed here
partial depinning because it is limited to the negative rolls, in contrast
with the depinning occurring globally for $v_0>v_{\rm th}$ \cite{Neufeld}.
This situation is graphically illustrated in Fig. \ref{F1}. For $\Omega_0 \ll
\Omega_L$, the particle's trajectories wander across the square array of
convection rolls undergoing marked changes of direction upon crossing them.
It is only for $\Omega_0 \sim \Omega_L$, that they start spiraling, but only
inside the positive subcells, panel (c). Finally, for $\Omega_0\gg \Omega_L$
trajectories appear to be the superposition of advection and counterclockwise
chiral rotations, the chiral rotations having much shorter a curvature radius
than the advection ones, panel (d). Therefore, a highly chiral particle is
{\it trapped} inside the convection rolls most of the time, irrespective of
their vorticity sign. As quantitatively confirmed by the numerical data
presented below, the characteristic self-polarization frequency, $\Omega_L$,
thus separates two distinct chirality regimes, respectively, of low,
$\Omega_0 \ll \Omega_L$, and high chirality, $\Omega_0 \gg \Omega_L$.

Based on this argument, we expect that the particle's stationary probability
density (pdf), $P(x,y)$, tends to accumulate inside the positive subcells. To
this purpose we integrated $P(x,y)$ over the positive and negative subcells,
separately, obtaining respectively the quantities $P_+$ and $P_-$ plotted in
Fig. \ref{F2}. Of course, being all pdf normalized to 1, in the absence of
chiral depinning, $P_+=P_-=1/4$. A strong spatial asymmetry of $P(x,y)$
emerges as $\Omega_0$ grows larger than $\Omega_L$.

\begin{figure}[bp]
\centering \includegraphics[width=8.5cm]{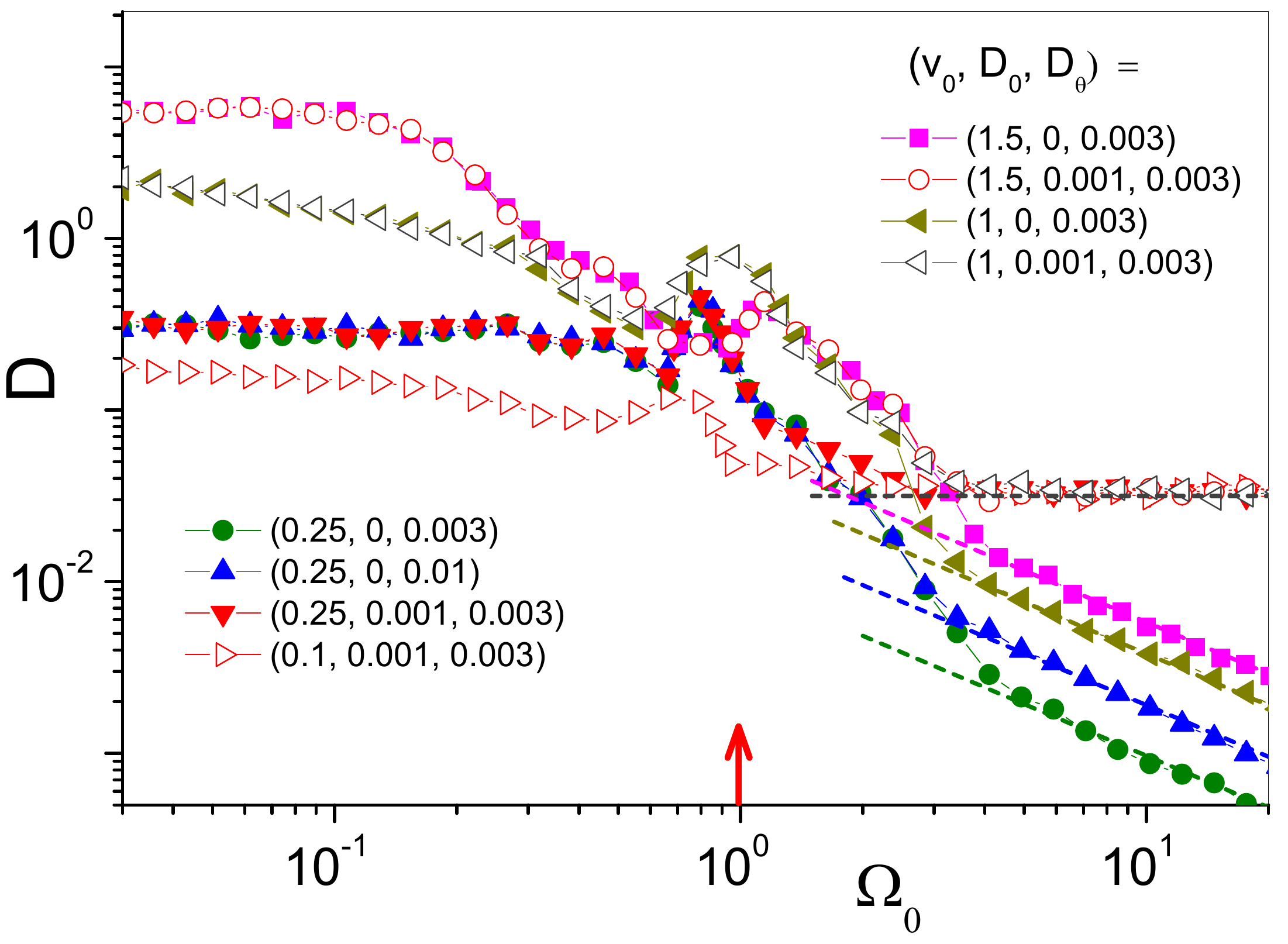}
\caption{Diffusion, $D$, vs. chiral torque, $\Omega_0$, in the cellular
flow  of Eq. (\ref{psi}) for different values of $D_0$, $D_\theta$ and $v_0$
(see legends). Other model parameters are: $\alpha=1$,  $U_0=1$, $L=2\pi$. Dashed horizontal
and sloped lines represent our analytical prediction, Eqs. (\ref{Dinfty})-(\ref{Dpomeau}).  The vertical arrow
corresponds to the condition $\Omega_0=\Omega_L$. \label{F3}}
\end{figure}

In the noiseless limit, $D_0=D_\theta=0$, depinning occurs slightly below the
self-polarization frequency, that is for $\Omega_0\simeq ~0.9 \Omega_L$. This
is consistent with the remark that $\Omega_L$ denotes the maximum vorticity
at the center of the $\psi(x,y)$ subcells. On the other hand, we also notice
that angular fluctuations with finite strength, $D_\theta >0$, no matter what
their physical origin, weaken the effect of the chiral torque, $\Omega_0$.
Consequently. in the presence of noise, the onset of partial depinning from
the negative $\psi(x,y)$ subcells gets ``delayed'', that is $P_{\pm}$ deviate
substantially from their uniform-distribution value, $1/4$, only for larger
$\Omega_0$ values, i.e., $\Omega_0>\Omega_L$.

\begin{figure}[bp]
\centering \includegraphics[width=8.5cm]{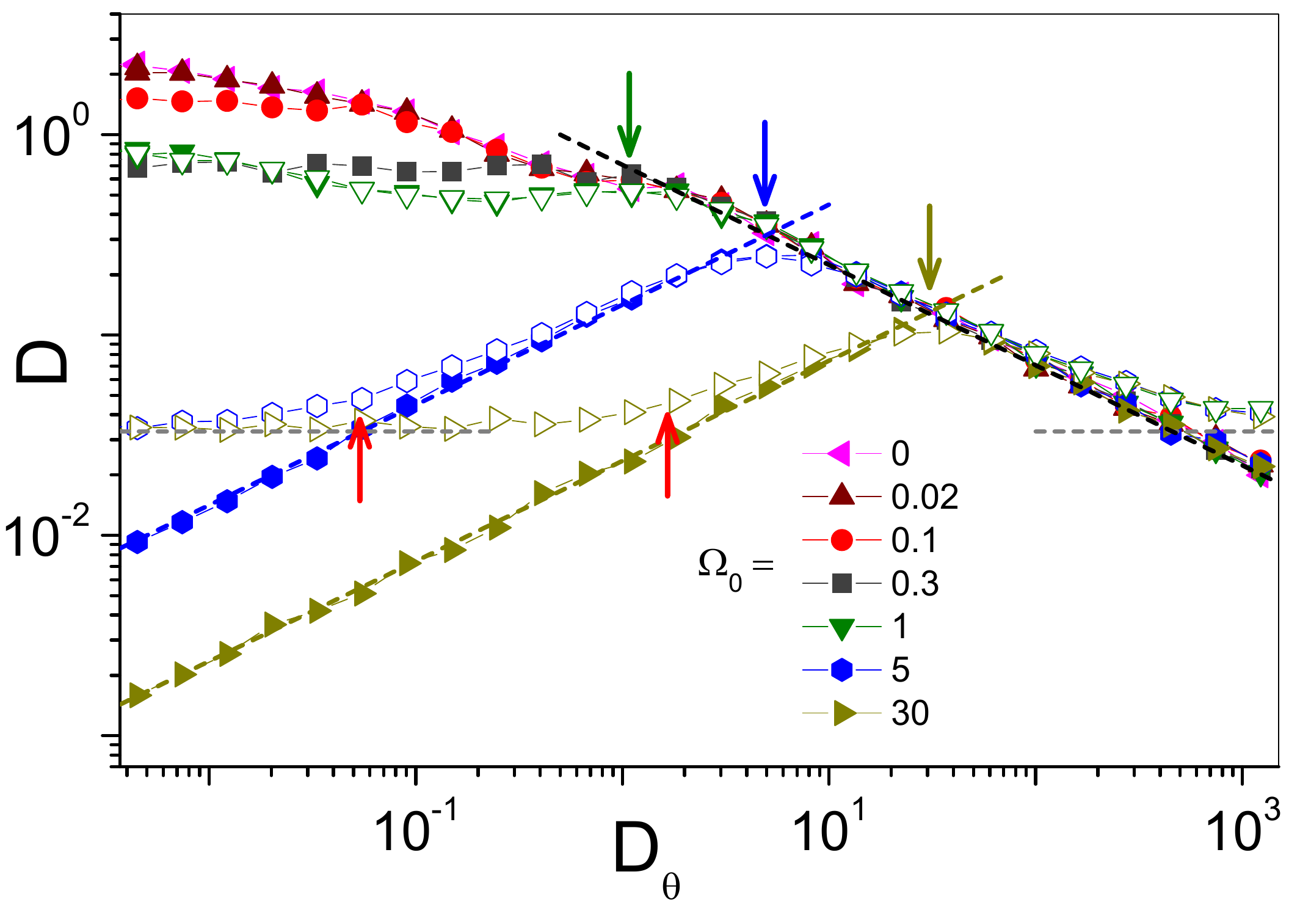}
\caption{Diffusion, $D$, vs. rotational noise strength, $D_\theta$, in the cellular
flow  of Eq. (\ref{psi}) for $D_0=0$ (solid symbols) and $0.001$
(empty symbols) and different $\Omega_0$  (see legend).
Other model parameters are: $\alpha=1$,  $U_0=1$, $L=2\pi$. Dashed lines represent
analytical predictions based on Eqs. (\ref{Dinfty}) and (\ref{Dpomeau}).
Downward and upward vertical arrows locate respectively the predicted maxima at
$D_\theta=\Omega_0$ and the emergence of the finite $D_0$  plateau at low $D_\theta$
(see text). \label{F4}}
\end{figure}

\subsection{Advection Controlled Diffusion} \label{advectivediff}

\begin{figure}[tp]
\centering \includegraphics[width=8.5cm]{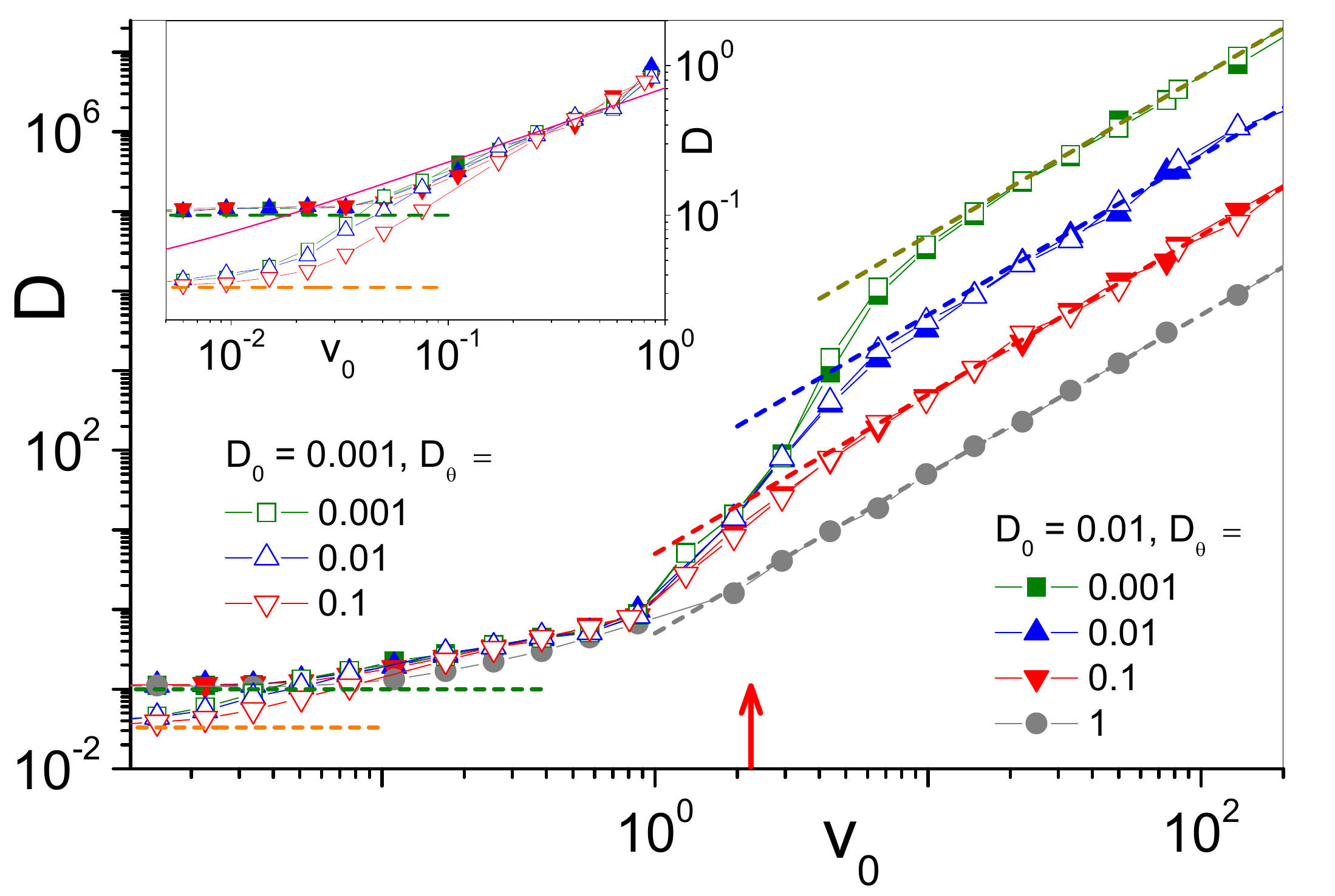}
\caption{Diffusion, $D$, vs. self-propulsion speed, $v_0$, in the cellular
flow  of Eq. (\ref{psi}) for $\Omega_0=0$ and different $D_0$ and $D_\theta$;
(see legend and inset for more details at low $v_0$).
Other model parameters are: $\alpha=1$,  $U_0=1$, $L=2\pi$.
Dashed horizontal
and sloped lines represent our analytical prediction (see text). The vertical arrow
corresponds to the condition $v_0=v_{\rm th}=2.2$ (see text) \cite{Neufeld}. \label{F5}}
\end{figure}

The strong chirality dependence of the trajectories shown in Figs.
\ref{F1}(b)-(d) and their nonuniform spatial localization illustrated in Fig.
\ref{F2}, have an immediate impact on the particle's asymptotic diffusion
constant, $D$. For a particle with $v_0/v_{\rm th} \ll \Omega_0/\Omega_L$ the
persistence length of its trajectories is much shorter than the flow cell
size, $v_0/\Omega_0 \ll L/2\pi$, so that its intracell diffusion constant is
well approximated by
\begin{equation}
\label{Dinfty}
\overline D_\infty= D_0+\frac{v_0^2}{2D_\theta}\frac{1}{1+(\Omega_0/D_\theta)^2},
\end{equation}
that is the diffusion constant of a chiral active particle in the absence
of advection \cite{Lowen,Ghosh2}.

We know \cite{Neufeld} that an achiral active particle, $\Omega_0=0$, with
$v_0 < v_{\rm th}$, crosses the subcell separatrices, Fig. \ref{F1}(b),  only
as an effect of its roto-translational fluctuations. On keeping ignoring
advection, its diffusion constant then would consist again of a translational
term, $D_0$, due to thermal noise and an additional term from {\it intercell
jumps} with effective step $L/2\pi$, namely \cite{Havlin}
\begin{equation}
\label{Dzero}
\overline D_0= D_0+D_L +  \frac{v_0}{2U_0}.
\end{equation}

We have now to take into account that diffusion occurs here in a flow pattern
of stream function $\psi(x,y)$. The effects of advection on the diffusion of
an active JP is illustrated in Fig. \ref{F3}. To interpret the numerical
results displayed there we notice that for $v_0 \ll U_0$ (trapped particle)
and $D_0 \ll D_L$ (weak noise-induced depinning, see Sec. III.C), both the
high- and low-chirality diffusion constants, $\overline D_\infty$, Eq.
(\ref{Dinfty}), and $\overline D_0$, Eq. (\ref{Dzero}), are much smaller than
$D_L$ (large effective P\'eclet number \cite{Pomeau}). This  suggests that
Eq. (38) of Ref. \cite{Pomeau} may apply to the case of active particles,
too. A simple extension of that equation to the stream function $\psi(x,y)$
of Eq. (\ref{psi}) yields the working fitting formula,
\begin{equation}
\label{Dpomeau}
D= (D_L \overline D)^{1/2},
\end{equation}
that is, for large P\'eclet numbers, the advective diffusion constant is
proportional to the square root of the no-flow particle's diffusion constant,
$\overline D$. The quantity $\overline D$ is approximated by Eqs.
(\ref{Dinfty}) and (\ref{Dzero}), respectively, in the high and low chirality
regimes.

In Fig. \ref{F3} the low and high chirality regimes are separated by an
excess diffusion peak centered at around $\Omega_0 \simeq \Omega_L$ (with a
weak dependence on $v_0$). This is the signature \cite{Costantini} of
particle's depinning from the negative subcells anticipated above. At high
$\Omega_0$ the data set plotted in Fig. \ref{F3} exhibit tails of two kinds,
depending on the value of $D_0$: (i) horizontal plateaus, $D= (D_L
D_0)^{1/2}$, insensitive to the self-propulsion parameters, for finite
thermal noise, and (ii) $D \simeq (D_\theta D_L/2)^{1/2}v_0/\Omega_0$ for
vanishingly translational noise, $D_0=0$. Both behaviors are closely
reproduced by Eq. (\ref{Dpomeau}) after replacing $\overline D$ with
$\overline D_\infty$ of Eq. (\ref{Dinfty}) (dashed lines).

The validity of Eq. (\ref{Dpomeau}) in the regime of high chirality,
$\Omega_0 \gg \Omega_L$, is also apparent in  Fig. \ref{F4}. For $D_0=0$, the
predicted diffusion constant $(D_L\overline D_\infty)^{1/2}$, Eqs.
(\ref{Dinfty})-(\ref{Dpomeau}), grows like $D_\theta^{1/2}$ for $D_\theta <
\Omega_0$, and then decays like $D_\theta^{-1/2}$ for $D_\theta > \Omega_0$,
after going through a maximum at $D_\theta=\Omega_0$ (downward arrows). Our
formula for $D$ fits closely the simulation data over the entire $D_\theta$
domain of Fig. \ref{F4}. For finite $D_0$, both the raising and decaying
branches are still visible, except they appear to merge into the thermal
plateau with $D=(D_L D_0)^{1/2}$.  This happens when the diffusion term due
to self-propulsion grows negligible with respect to $D_0$. Accordingly, for
instance, at low $D_\theta$ the thermal plateau extends up to $D_\theta \sim
2(\Omega_0/v_0)^2 D_0$ (upward arrows).

In the low chirality regime, $\Omega_0 < \Omega_L$,  the raising branch of
the $D$ curves of Fig. \ref{F4} is replaced by a horizontal branch, which
weakly depends on the angular frequencies, $\Omega_0$ and $D_\theta$.
Moreover, the horizontal and the decaying branches of the low chirality $D$
curves connect around $D_\theta \sim \Omega_L$. Indeed, when the chiral
frequency, $\Omega_0$, is lowered below the self-polarization frequency,
 the natural frequency for the angular rate, $D_\theta$, to
compare with is now $\Omega_L$.
For $\Omega_0 > {\rm max}\{\Omega_L, D_\theta\}$ the JP behaves as a regular
Brownian particle with effective local diffusion constant,
$D_0+v_0^2/2D_\theta$, see Eq. (\ref{Dinfty}); the $D$ decaying branch is
therefore the same for both low and high chirality.

Such a distinct diffusion regime is better illustrated in Fig. \ref{F5},
where we study the constant $D$ as a function of the self-propulsion speed,
$v_0$. As expected, for $v_0 \gg v_{\rm th}$, the particle is largely
insensitive to the advective drag, so that its diffusion constant approaches
the zero-flow value, $\overline D_\infty$, of Eq. (\ref{Dinfty}), i.e., $D$
is quadratic in $v_0$ (sloped dashed lines). The curves plotted in the main
panel of Fig. \ref{F5} show a sharp jump in the vicinity of the threshold,
$v_{\rm th}$, thus confirming the existence of the depinning mechanism
introduced in the noiseless limit by the authors of Ref. \cite{Neufeld}.
Relevant to the present study is the $v_0$ dependence of $D$ below the
depinning threshold, $v_0<v_{\rm th}$.  As anticipated above, we expect that
formula (\ref{Dpomeau}) applies to a flow trapped particle also at low
chirality, $\Omega_0 < \Omega_L$, provided that $\overline D$ is replaced by
$\overline D_0$ of Eq. (\ref{Dzero}). For vanishing values of $v_0$, we
recover the expected limit $(D_0 D_L)^{1/2}$ (horizontal dashed lines),
whereas for $D_0/D_L \ll v_0/U_0 \ll 1$ the constant $D$ grows proportional
to $v_0^{1/2}$ (inset of Fig. \ref{F5}). We remind here that in the noiseless
limit, $D_0, D_\theta \to 0$, the particle's dynamics becomes chaotic
\cite{Neufeld} and diffusion is suppressed (and hard to compute numerically).

\begin{figure}[tp]
\centering \includegraphics[width=8.5cm]{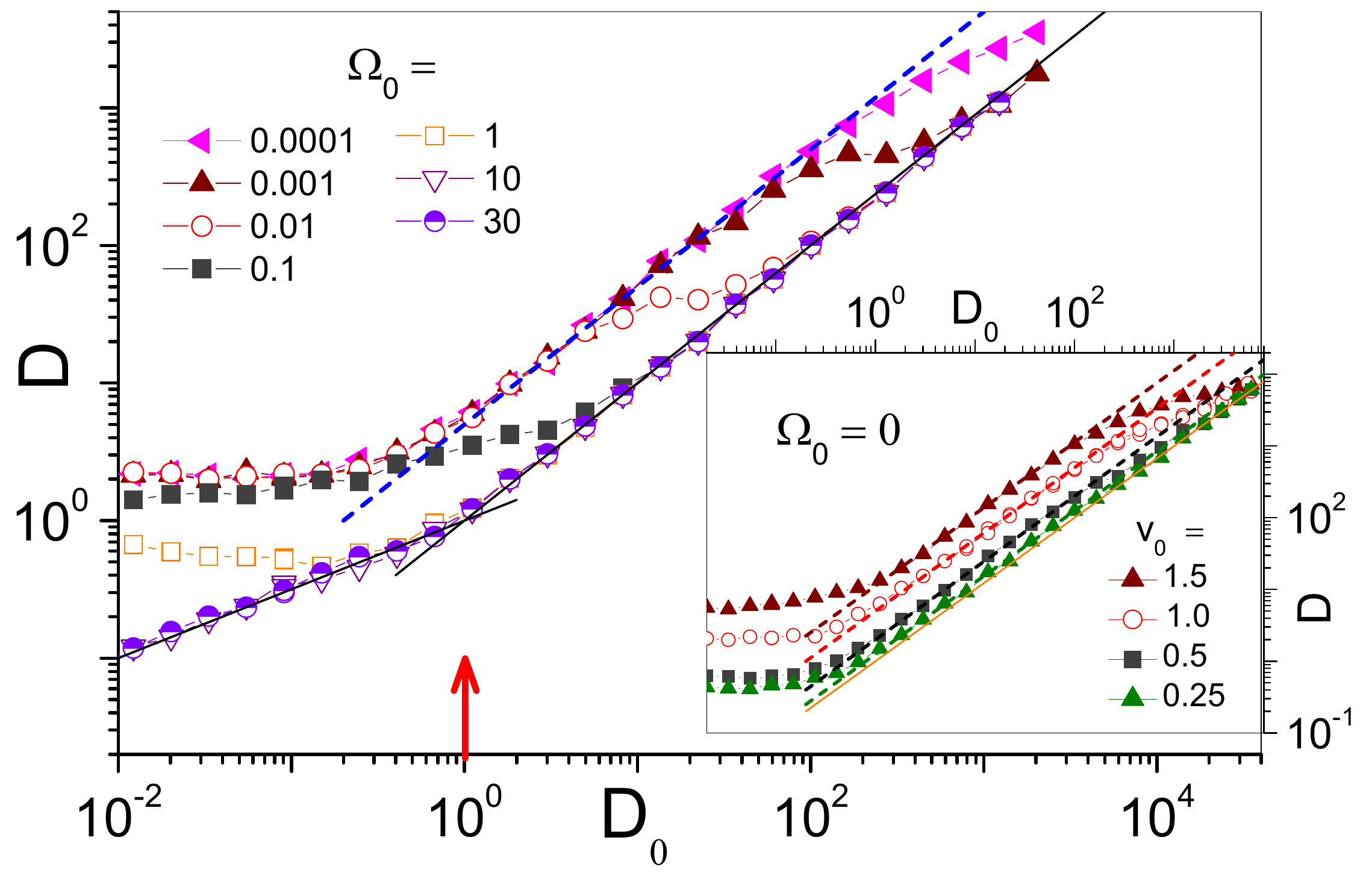}
\caption{Diffusion, $D$, vs. translational noise, $D_0$, in the cellular
flow  of Eq. (\ref{psi}) for $D_\theta=0$ and main panel: $v_0=1$ and different
$\Omega_0$, inset:  $\Omega_0=0$ and different $v_0$
(see legends). Other model parameters are: $\alpha=1$,  $U_0=1$, $L=2\pi$. Straight
lines represent the analytical predictions discussed in the text for high
(solid) and low chirality (dashed). A vertical arrow locates $D_0=D_L$.  \label{F6}}
\end{figure}

\subsection{Noise Controlled Diffusion} \label{therdiff}

The effects discussed in the foregoing subsections are detectable only at low
thermal noise levels. In Fig. \ref{F6} we illustrate how advection effects
can be washed out by large thermal noise, even under the depinning threshold,
i.e., for $v_0<v_{\rm th}$.  In the case of a high chirality particle,
$\Omega_0 \gg \Omega_L$, at low noise, $D_0 \ll D_L$, we know that $D=(D_L
D_0)^{1/2}$, whereas for exceedingly large $D_0$ we expect $D=D_0$ (free
diffusion limit). The transition between these two limits would take place
for $D_0 \sim D_L$. This is consistent with the simulation data of Fig.
\ref{F6} (vertical arrow). The case of low chirality is more interesting. The
large noise branch of $D$ till sets out proportional to $D_0$, but with
substantially larger slope, which seems to increase proportionally to $v_0^2$
(see inset). Our numerical data for $D$ finally approach the free diffusion
law, $D=D_0$, but asymptotically, only, around $D_0 \sim U_0^2/\Omega_0$. To
explain this phenomenon we remark that at low P\'eclet numbers, $D_0\gg D_L$,
the active particle is no longer trapped in the convection rolls. For low
angular rates, $D_\theta, \Omega_0 \ll \Omega_L$, its mean free path is of
the order of $v_0 \tau_L$, where $\tau_L=L/2\pi U_0$ is the effective cell
crossing time, and gets scattered against the cell separatrix with (short)
persistence time, $\tau_0=(L/2\pi)^2/8D_0$, governed by the translational
noise. Using the argument invoked to derive Eq. (\ref{Dzero}), we predict,
$D=D_0[1+4(v_0/U_0)^2]$, in good agreement with the simulation data reported
in Fig. \ref{F6}. By the same token, one locates the switching between such a
transient law and the free diffusion law at around $D_0 \sim
U_0^2/8\Omega_0$.

In conclusion, our simulations prove that the combination of advection and
self-propulsion determines an appreciable excess diffusion of weakly chiral
active JPs even at low P\'eclet numbers.

\section{Conclusions} \label{conclusions}

In this paper we have shown how active particles in hydrodynamically active
mediums exhibit peculiar diffusion properties, which distinguish them from
common colloidal particles. This is particularly true in the low chirality
regime, where self-propulsion determines a rich phenomenology of the
diffusion process. We remind that the simple and best known stream function
$\psi(x,y)$ of Eq. (\ref{LE}), models situations that have been already
implemented experimentally, e.g., with rotating cylinders \cite{Gollub2} or
with ion solutions in arrays of magnets \cite{Tabeling}. Moreover, the
numerical and analytical techniques reported here can be easily extended to
different stream functions to represent convection rolls of varying
topologies \cite{Crisanti}. It is clear from this investigation that, in view
of technological applications, advection controlled diffusion should be
considered as an effective tool to govern the transport of active matter.
Important examples are microfluidic devices \cite{Squires} or even
microswimmer diffusion in steady turbulent flows \cite{Tabeling}.

\section*{Acknowledgements}
Y.L. is supported by the NSF China under grants No. 11875201 and No.
11935010. P.K.G. is supported by SERB Start-up Research Grant (Young
Scientist) No. YSS/2014/000853 and the UGC-BSR Start-Up Grant No.
F.30-92/2015. D.D. thanks CSIR, New Delhi, India, for support through a
Junior Research Fellowship.

\end{document}